\documentclass{article}

\usepackage{amsmath}
\usepackage{lineno} 
\usepackage{graphicx}
\modulolinenumbers[5]

\title{Transformation Optics for Plasmonics: from Metasurfaces to Excitonic Strong Coupling}
\author{Paloma A. Huidobro$^{1,*}$, Antonio I. Fern\'andez-Dom\'{i}nguez$^{2,\dagger}$ }

\date{%
	\small{
    $^1$Instituto de Telecomunicações, Insituto Superior Te\'ecnico-University of Lisbon, Avenida Rovisco Pais 1,1049-001 Lisboa, Portugal\\%
    $^2$Departamento de F\'{i}sica Te\'orica de la Materia Condensada and Condensed Matter Physics Center (IFIMAC), Universidad Autónoma de Madrid, E-28049 Madrid, Spain\\
    $^*$p.arroyo-huidobro@it.lx.pt  \\
    $^\dagger$a.fernandez-dominguez@uam.es 
    } \\ [2ex]%
    \today
}

\begin{document}



\maketitle 


%
%

\begin{abstract}
We review the latest theoretical advances in the application of the framework of Transformation Optics for the analytical description of  deeply sub-wavelength electromagnetic phenomena. First, we present a general description of the technique, together with its usual exploitation for metamaterial conception and optimization in different areas of wave physics. Next, we discuss in detail the design of plasmonic metasurfaces, including the description of singular geometries which allow for broadband absorption in ultrathin platforms. Finally, we discuss the quasi-analytical treatment of plasmon-exciton strong coupling in nanocavities at the single emitter level.
\end{abstract}




\section{Introduction}
\label{sec:Intro}
The development of Transformation Optics~\cite{Ward1996,Pendry2006} (TO) has been instrumental in the fast development that metamaterial science has experienced during the last years~\cite{Chen2010}. This theoretical tool exploits the invariance of macroscopic Maxwell's equations under coordinate transformations to establish a link between an electromagnetic (EM) phenomenon, described by the transformation, and the material response required for its realization. Thus, TO determines the way in which the EM constitutive relations, and therefore the permittivity and permeability tensors, must be tailored in space in order to obtain a desired effect. 

\begin{figure}[!t]
\includegraphics[width=1\linewidth]{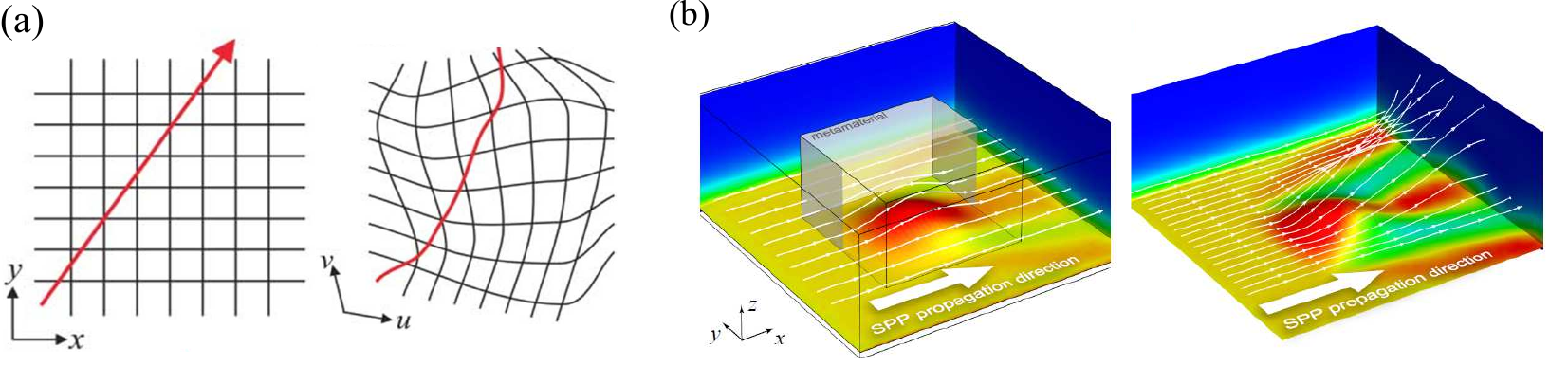}
\vspace{-0.25 cm} \caption{(a) EM fields propagation in free space
(sketched as a single field line) with the background cartesian
grid (top), and their distortion under an arbitrary geometric
transformation, with the corresponding distorted coordinates in
the background (bottom). Adapted with permission~\cite{Pendry2006}. (b) A bump on a metal surface can be cloaked so that a SP propagating along the metal/dielectric interface propagates smoothly without being scattered as would occur without the cloak (right panel). Powerflow streamlines are depicted as with lines with arrows. Reproduced with permission from Ref. \cite{Huidobro2010}  }
\label{fig:Intro}
\end{figure}

TO theory states that, under a general spatial transformation, ${\bf r}'={\bf r}'({\bf r})$ like the one sketched in Figure~\ref{fig:Intro}(a), EM fields are modified exactly in the same way as they do for the following spatially-dependent electric
permittivity and magnetic permittivity tensors
\begin{equation}
\boldsymbol{\epsilon}'({\bf r}') = \frac{\boldsymbol{\Lambda}({\bf r}')\boldsymbol{\epsilon}({\bf r}({\bf r}'))\boldsymbol{\Lambda({\bf r}')}}{\sqrt{{\rm det}[\boldsymbol{\Lambda}({\bf r}')]}},\,\,
\boldsymbol{\mu}'({\bf
r}') = \frac{\boldsymbol{\Lambda}({\bf r}')\boldsymbol{\mu}({\bf
r}({\bf r}'))\boldsymbol{\Lambda({\bf r}')}}{\sqrt{{\rm
det}[\boldsymbol{\Lambda}({\bf r}')]}} .\label{eq:trans}
\end{equation}
where $\boldsymbol{\epsilon}({\bf r})$ [$\boldsymbol{\mu}({\bf
r})$] and $\boldsymbol{\epsilon}'({\bf r}')$
[$\boldsymbol{\mu}'({\bf r}')$] are the permittivity
[permeability] tensors in the original and final frames,
respectively, and $\Lambda({\bf r}')=\partial{\bf r}'/\partial{\bf
r}$ is the Jacobian matrix for the transformation.

From a metamaterial perspective, Equations~\ref{eq:trans}
establish the link between material characteristics and the EM effect resulting from the spatial operation. Thus, TO provides a recipe for the design of metamaterials with at-will functionalities. A recent review on the use of transformation optics for the design of cloaks, illusion devices and other elements such as rotators and concentrators can be found in Ref. \cite{sun2017transformation}. In parallel to the development of optical metamaterials for such applications, there has been a wide variety of advances in different areas of wave physics. For instance, TO has been extended into the spatiotemporal domain to devise spacetime cloaks \cite{mccall2010spacetime,fridman2012demonstration} as well as analogues of other phenomena emerging from the link between electromagnetism in media and general relativity \cite{leonhardt2006general,greenleaf2007electromagnetic,smolyaninov2010metric,thompson2010completely}. Another degree of design flexibility in the form of media with gain and loss can be obtained by analytic continuation of the mapping coordinates into the complex plane \cite{horsley2016wave,horsley2015spatial}. This way, TO can be connected with PT symmetric media \cite{ruter2010observation}, and reflectionless devices can be designed \cite{castaldi2013p}. TO has also been adapted for the control of surface waves  \cite{mitchell2013perfect}, as well as guided waves on integrated optical circuits \cite{Viaene2016}, and for antenna engineering \cite{quevedo2013transformation}. 

Furthermore, it is possible to apply TO to other wave systems beyond electromagnetics. The cornerstone proposal of the invisibility cloak \cite{Schurig2006} has been reproduced in different fields, which initiated the expansion of metamaterials for different physical domains. Applications of TO include the cloaking of acoustic pressure waves \cite{cummer2007one,chen2007acoustic,Zigoneanu2014}, matter waves \cite{zhang2008cloaking} or heat conduction \cite{guenneau2012transformation}. Particularly interesting is the fact that while the elastodynamic equations are not form invariant under coordinate transformations \cite{Milton2006}, it has been shown that TO can still be a powerful tool in the design of mechanical metamaterials \cite{stenger2012experiments,buckmann2014elasto,buckmann2015mechanical} and even of seismic cloaks \cite{Brun2009,Guenneau2014}. 

Much the development of transformation optics has gone hand in hand with that of plasmonics. 
In its most general form, TO accounts for the vectorial and undulatory nature of EM fields, which makes it exact at sub- and supra-wavelength scales. 
Taking advantage of this, TO has made possible the at-will moulding the flow of surface plasmons (SPs) that propagate along metal/dielectric interfaces with subwavelength confinement \cite{Huidobro2010,liu2010transformational,Kadic2012}. Although the SP field extends both into the dielectric and metallic sides of the interface, it has been shown that it is enough to act on the dielectric side by placing the metamaterial designed with Eqs. \ref{eq:trans} on top of the metal surface \cite{Huidobro2011}. This way, plasmonic invisibility cloaks, such as shown in Fig.~\ref{fig:Intro} (b), beam benders and shifters operating at nearly subwavelength scales and in the visible regime have been devised. Experimental realizations of these ideas include a broadband carpet cloak which suppresses scattering from a bump on a metal surface \cite{Renger2010}, as well as Luneburg and Eaton lenses \cite{zentgraf2011plasmonic}. 

From a purely computational electrodynamics perspective,
Equations~\ref{eq:trans} provide the prescription to interchange
geometric and material characteristics of an EM system. This was,
in fact, the original motivation that led to the development of
this theoretical framework. It was devised as a strategy to ease the numerical
solution of Maxwell's Equations, using the TO mapping of
complex and acute geometries into much simpler
ones~\cite{Ward1996}. Frequently, this advantage comes at the
expense of non-uniform and anisotropic permittivity and
permeability distributions. This initial purpose of TO has been
recovered in recent years. It has been used as a means to shed
analytical, instead of numerical, insight into plasmonic phenomena
taking place in deeply subwavelength metallic devices.

At visible frequencies and (sub-)nanometric length scales, spatial
derivatives in Maxwell's curl Equations are much larger than
temporal ones. Therefore, the latter can be neglected, which
translates into the decoupling of magnetic and electric fields.
This is the so-called quasistatic
approximation for metallic nanostructures~\cite{novotny2012principles}, in which the spatial dependence of electric
fields can be described in terms of an electrostatic potential,
${\bf E}({\bf r})=-\nabla\Phi({\bf r})$ , satisfying Gauss law
\begin{equation}
\nabla[\boldsymbol{\epsilon}({\bf r})\nabla\Phi({\bf
r})]=0,\label{eq:Gauss}
\end{equation}
where, in general, the permittivity is an inhomogeneous,
anisotropic tensor. Importantly, although the quasistatic
approximation only holds for sub-wavelength systems, the validity
of Equation~\ref{eq:Gauss} can be pushed to dimensions up to a
${\sim100}$~nm by introducing radiation losses through the
so-called radiative reaction
concept~\cite{novotny2012principles,aubry2010conformal}. Using
these ideas~\cite{pendry2012transformation,luo2013harvesting,pendry2015transforming}, a set of analytical
and quasianalytical TO approaches have
been devised to investigate the harvesting of light by a wide range of 2D and 3D geometries: touching
nanoparticles~\cite{aubry2010plasmonic,fernandez2010collection},
nanocrescents~\cite{aubry2010broadband,fernandez2012theory},  nanorods~\cite{kraft2014transformation},
nanosphere dimers~\cite{aubry2011plasmonic,pendry2013capturing}. Moreover, other EM phenomena have been explored theoretically using TO ideas, such as spatial nonlocality in metallic junctions~\cite{fernandez2012transformation,fernandez2012insight}, electron energy loss in metal nanostructures~\cite{kraft2016transformation}, second harmonic generation in plasmonic dimers~\cite{reddy2019surface}, near-field
van der Waals interactions between nanoparticles~\cite{zhao2013description,luo2014van}, or plasmon hybridization in collections of several touching nanoparticles \cite{YuPNAS2019}. 

In the following, we discuss the recent exploitation of TO framework in two  areas of great interest in plasmonics in recent years. On the one hand, the design of conventional and singular plasmonic metasurfaces, which can be metallic or based on graphene. On the other hand, the description of strong-coupling phenomena between quantum emitters and the plasmonic spectrum supported by metallic nanocavities. 

\section{Plasmonic metasurfaces}

Metasurfaces, the planar counterpart of bulk metamaterials, consist of resonant subwavelength units arranged in a two-dimensional (2D) array \cite{holloway2012overview,kildishev2013planar,Minovich2015,GLYBOVSKI20161,Monticone2017,SpoofBook}. The geometry and materials of the subwavelength building blocks, as well as their arrangement, are appropriately designed and manufactured to provide an ultra-thin platform for manipulating EM waves. Metasurfaces have enabled effects such as broadband light bending and anomalous reflection and refraction in ultrathin platforms \cite{Ni2011,Yu2011}. While dielectric nanoantennas have been suggested for the design of metamaterials due to their lower loss compared to plasmonic nanoparticles \cite{Genevet2017recent}, absorption losses are a less stringent constrain when considering metasurfaces. For that reason, plasmonic metasurfaces have been a particularly fruitful platform to control optical fields \cite{meinzer2014plasmonic}. They are formed of subwavelength metallic elements with resonant electric or magnetic polarizabilities, enabling light confinement at the subwavelength scale, accompannied by large enhancements of the EM fields. On the other hand, the high electron mobility in graphene has also motivated the use of this 2D material for plasmonic metasurfaces at lower frequencies, making use of the unrivalled field enhancements provided by its THz plasmons \cite{Koppens2011,Ju2011,Nikitin2011,Grigorenko2012,Low2014}. 

The analytical power of TO has been instrumental in the  design of plasmonic metasurfaces with unconventional properties, as we review in the following. In Section \ref{sec:gratings}, we discuss in detail the TO insights into 
both subwavelength metallic gratings and graphene metasurfaces, as well as their applications. Next, in Section \ref{secC:singularmetasurfaces} we move on to present the so-called singular metasurfaces, their fundamental properties and their understanding in terms of hidden dimensions.

\subsection{Designing plasmonic gratings with transformation optics}
\label{sec:gratings}

\begin{figure}[t!]
    \begin{center}
        \includegraphics[width=\textwidth]{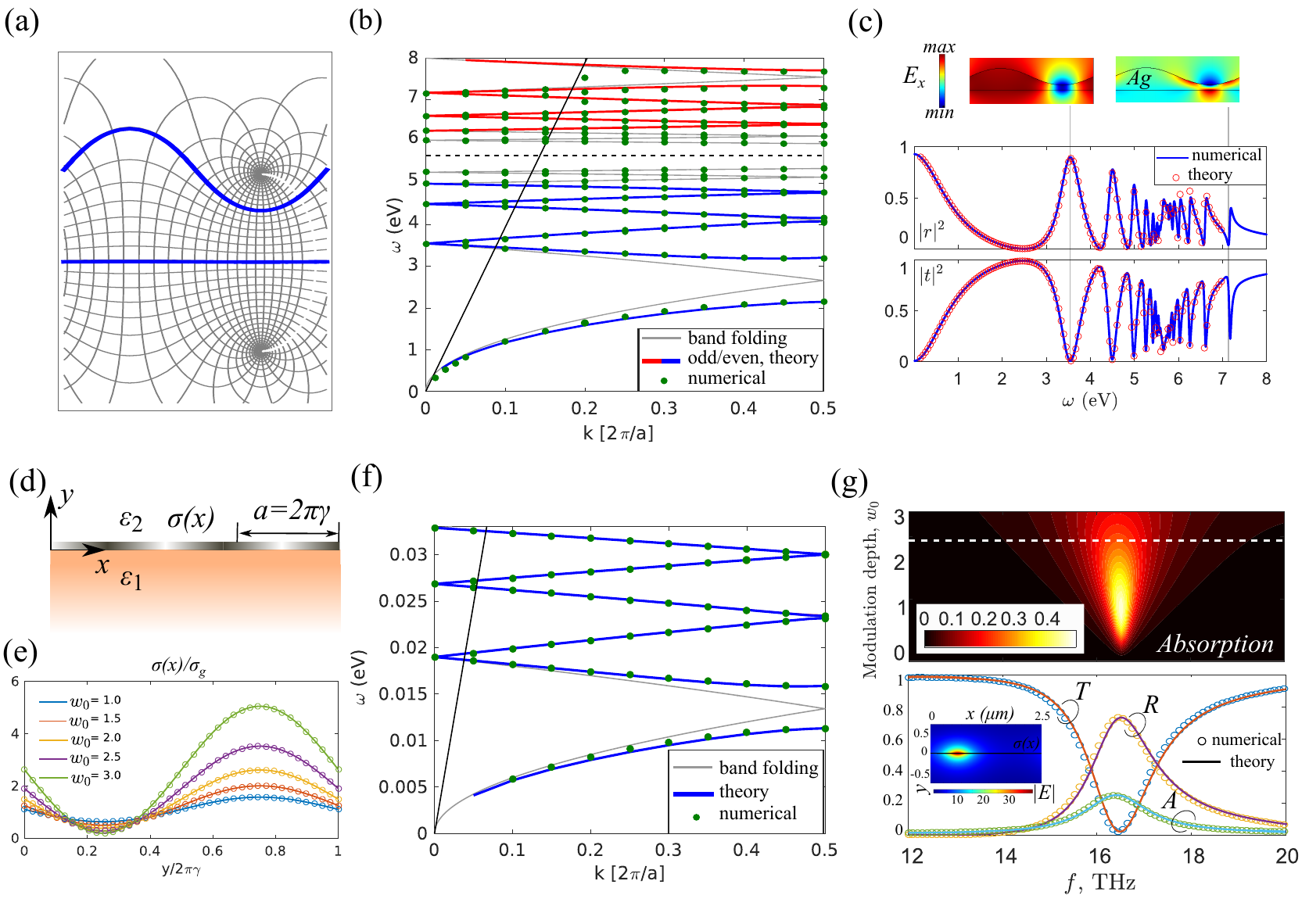} 
        \caption{\label{fig:gratings} Realising plasmonic metasurfaces by means of conformal transformations. (a) One period of the conformal map used to generate gratings. (b) Dispersion relation of a silver grating modes above and below the SP frequency (dashed line). The grating inherits the mode spectrum of the flat slab (thin gray line). (c) Optical response of the subwavelength silver grating at normal incidence, and field distribution for the lowest and highest energy modes. (d) Graphene metasurfaces with spatially varying conductivity, $\sigma(x)$, and tunable modulation strength (e). (f) Dispersion relation of a graphene metasurface revealing the underlying homogeneous graphene layer (thin gray line). (g) Absorption can reach 50$\%$ at the dipolar SP resonance by tuning the modulation strenght (top). Absorption, reflection and transmission for $w_0=2.5$ (bottom). Panels (b,f) are adapted from Ref. \cite{Huidobro2017a} , (c) from \cite{Kraft2015}, (d,e) from \cite{Huidobro2016a}.}
    \end{center}
\end{figure}

Here we review the theoretical framework for the design of metasurfaces by means of TO. We concentrate on the most simple form of plasmonic metasurfaces, that is, a thin film of a plasmonic material one of whose surfaces is periodically corrugated forming a subwavelength  grating \cite{Chandezon1980,barnes1996}. 
Such plasmonic grating can be generated from a thin metallic slab (where analytical solutions of Laplace's equation are available) by means of a conformal transformation \cite{Kraft2015}, 
\begin{equation}
    z = \frac{d'}{2\pi} \ln \left(\frac{1}{e^w-iw_0} +iy_0 \right).
\end{equation}
Here, $z = x+iy$ refers to the transformed coordinates in the frame of the grating, and $w = u+iv$ to the Cartesian coordinates in the frame of the slab. In addition, $ d' $ sets the length scale of the structure by determining the grating period, $ w_0$ is a free parameter that sets the grating modulation strength, and $y_0$ is fixed by $w_0$, the slab thickness, $\delta$, and its position, $u_0$, as $y_0 = w_0/( \exp [2(u_0 + \delta)]- w^2_0)$. A map of one period of the conformal transformation is shown in Fig. \ref{fig:gratings} (a). The space between the blue lines represents a silver slab with one periodically modulated surface, which maps through the transformation to a flat silver slab.

As we have discussed in Section  \ref{sec:Intro}, conformal transformations applied to Maxwell's equations conserve the electrostatic potential. Hence, in the electrostatic limit (period $d' \ll \lambda$), the spectral properties of a subwavelength plasmonic grating are equivalent to that of the thin plasmonic slab, whose dispersion relation is given by $ \exp(|k|\delta) =  \pm ( \epsilon_m( \omega)-\epsilon_d)/(\epsilon_m( \omega)+\epsilon_d)$, $\epsilon_m=1-\omega_p^2/(\omega(\omega+i\gamma_D))$ being the metal permittivity ($\omega_p$ is the plasma frequency and $\gamma_D$ is the Drude damping) and $\epsilon_d$ the one of the surrounding dielectric space. As a consequence, the dispersion relation of the grating can be accurately predicted from the simple analytical expression corresponding to the plasmonic slab. 

Figure \ref{fig:gratings}(b) shows the dispersion relation of a vacuum-embedded silver slab folded in the first Brillouin zone of the corresponding set of gratings (periodicity $ 2\pi \times 5 $ nm), plotted as a thin gray line. The modes of a grating with modulation strength fixed by $w_0 = 1.5$ calculated from a numerical finite element solver (comsol Multiphysics) are plotted with dots, presenting an excellent agreement with the analytical prediction. The relationship between both systems ensures the quasi-degeneracies observed at the zone center, which are only slightly lifted due to magnetic effects (the magnetic field sees the periodicity of the grating through variations in the out-of-plane component of the permeability). On the other hand, band gaps open at the zone edge. This reflects the periodic character of the grating  structure, since the slab is translationally invariant and only the modes at $k=0$ share exactly the same symmetry. Mathematically this is captured by the branch cuts of the transformation, which act as sinks and drains for the waves, effectively swapping them to opposite sides of the slab when incident and reflected waves on the grating are transformed. By taking this into account, band structures for the whole Brillouin zone, exact in the quasistatic limit, can be obtained \cite{Huidobro2017a} (plotted as thick red and blue lines). Furthermore, by going beyond the quasistatic approximation in a perturbative approach and including the radiative contribution of the grating, the optical spectrum at normal incidence can be obtained analytically [shown in Fig. \ref{fig:gratings}(c)].  Finally, we remark that the general scope of TO has enabled to fully take into account retardation effects by transforming the full set of Maxwell's equations. This enables the semi-analytical calculation of optical spectra for gratings of periods not limited to the very subwavelength regime, for arbitrary polarization states, and is exact at the level of Maxwell's equations \cite{PendryPhysRevB2019}.

The conformal map shown in Fig. \ref{fig:gratings}(a) can also be used to devise graphene metasurfaces, see panel (d). We consider the limit of an infinitely thin plasmonic slab with conductivity $\sigma(\omega)=-i(\epsilon(\omega)-1)\omega\epsilon_0 \delta $ and $\delta\rightarrow 0$. In the grating frame, the slab of modulated thickness equivalently represents an infinitely thin layer, i.e., graphene, with modulated conductivity \cite{huidobro2016graphene}. Metasurfaces consisting of graphene with periodically modulated conductivity \cite{peres2013exact,slipchenko2013,chen2017flatland} can be designed this way \cite{Huidobro2016a}, and a periodic doping modulation can be realized by optical~\cite{baudisch2018ultrafast} or electrostatic~\cite{bokdam2011electrostatic}  means, or by patterning the graphene~\cite{Fan2015,Nikitin2012b} or its environment~\cite{vakil2011,iranzo2018probing}. 

The dispersion relation of a graphene metasurface is shown in panel (b), displaying the  quasi-degeneracies at the zone center inherited from the dispersion relation of homogenously doped graphene (thin gray line).  The modulation period is $2.5$ $\mu$m, the modulation strength is given by $\omega_0=1.5$, see panel (e), and graphene's conductivity is taken from the random phase approximation with chemical potential $\mu=0.1$ eV and scattering loss $\tau = 10 $ ps. A close up of the absorption spectrum around the dipolar resonance, lower energy mode in panel (f), is presented in panel (g). Here the chemical potential was changed to $\mu=0.65$ eV, which accounts for the frequency shift with respect to the resonance in panel (f) and a typical experimental mobility of $10^4$ cm$^2/$(V$\cdot$s) was used. The insensitivity of the absorption peak in the contour plot to the modulation strength, $w_0$, is due to the fact that gratings of different $w_0$ map into homogeneous graphene with the same conductivity, as this is a free parameter in the transformation. Hence, by tuning the modulation depth, absorption in the graphene metasurface can be switched, and, remarkably, up to $50\% $ of the power of incident radiation can be absorbed by a single graphene layer owing to the excitation of deeply subwavelength  SPs. While $50\% $ absorption is the theoretical maximum for a thin layer of material, absorption can be further increased up to  $100\% $ by employing a Salisbury screen scheme and placing the metasurface close to a perfect reflector, such that a Fabry-Perot cavity is formed. Due to the strong EM confinement enabled by SPs, this idea allows for an ultrathin perfect absorber of deeply subwavelength thickness for THz frequencies \cite{Huidobro2017}.

\subsection{Singular plasmonic metasurfaces}
\label{secC:singularmetasurfaces}

As discussed in Section \ref{sec:Intro}, TO provided a successful understanding of the harvesting of light by plasmonic nanoparticles with singular geometries such as touching points \cite{aubry2010plasmonic,fernandez2010collection}. In particular, TO highlighted the geometrical origin of the broadband absorption spectra characteristic of these systems by mapping them to infinitely extended geometries where the singularites map into points at infinity. The infinite extension in the transformed frame removes the quantization (discretization) condition and yields a broadband spectrum, while the large absorption efficiencies are caused by the SP fields travelling towards infinity with reducing group velocities. This effect has also been referred to as anomalous absorption as it is present even in the absence of material loss \cite{Alu2013unbounded,Wallen2015anomalous}. 

\begin{figure}[t!]
    \begin{center}
        \includegraphics[width=\textwidth]{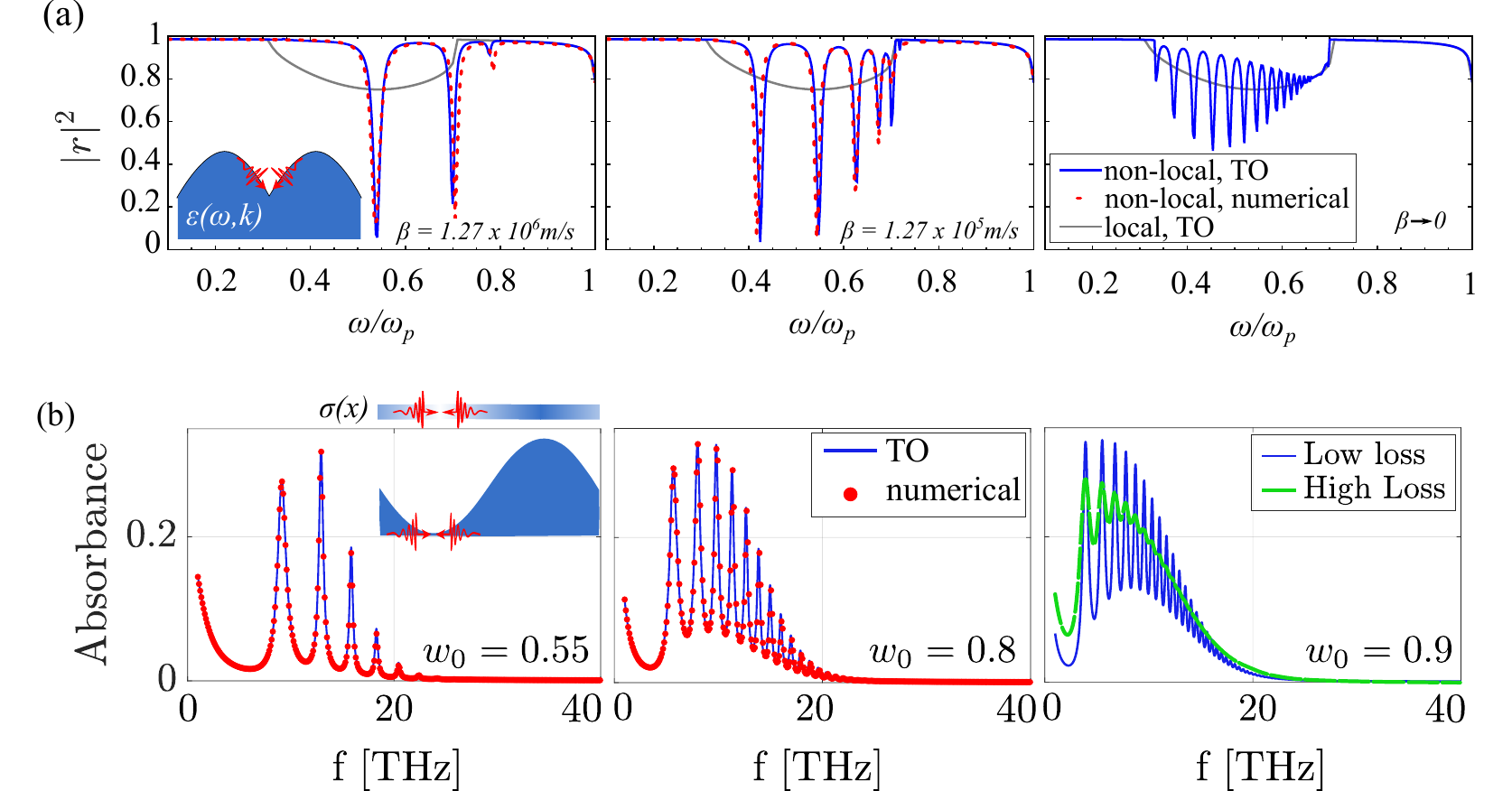}
        \caption{\label{fig:singulargratings} Optical response of singular metal (a) and graphene (b) metasurfaces. (a)  Reflectivity spectrum of a silver surface with a periodic array of sharp grooves. As $\beta$ is decreased from the realistic value towards zero (left to right), the spectrum  approaches the continuum predicted by a local calculation. (b) Absorption spectrum for graphene metasurfaces as the singular case is approached by reducing $\sigma(\omega)$ at the grating valleys ($w_0$ increasing, left to right). In both panels TO analytical results are plotted as solid lines and EM simulations are plotted with dots. Adapted from Refs. \cite{YangPhysRevB2019} (a) and \cite{galiffi2018broadband} (b). }
    \end{center}
\end{figure}

The physics of SPs propagating towards geometrical singularities can also play a fundamental role in plasmonic metasurfaces. For instance, subwavelength arrays of ultra sharp grooves in a metal surface can turn good reflectors such as gold or silver into almost perfect broadband absorbers, usually referred to as plasmonic black metals \cite{sondergaard2012plasmonic}. These surfaces can be viewed as \textit{singular plasmonic metasurfaces} \cite{Pendry2017} and here we review two instances of them, sheding light onto their continuous absorption spectra and highlighting how these can be interpreted by means of an extra compacted dimension. In the first case the singular metasurface consists of a periodic array of grooves with sharp edges carved on a metal surface \cite{Yang2018,YangPhysRevB2019}, see inset of Fig. \ref{fig:singulargratings}(a). In the second case, the singularities are achieved by strongly supressing the conductivity of graphene at the grating valleys \cite{galiffi2018broadband,Galiffi2019epj}, see inset of Fig. \ref{fig:singulargratings}(b). 

Plasmonic black metals \cite{sondergaard2012plasmonic} can be analytically modelled using TO by means of the periodic surface with sharp grooves shown in the inset of Fig. \ref{fig:singulargratings} (a) \cite{Pendry2017}. The sketch shows one period of the structure, which results from first mapping a set of plasmonic slabs that are infinite in the horizontal direction and periodically arranged in the vertical direction into a knife edge, compressing $-\infty$ to a point at the origin. Then $+\infty$ is compressed into another point, giving rise to a finite lenticular shape with two sharp edges. Finally, a logarithmic transformation is used to generate a semi-infinite surface decorated with a periodic array of grooves [see panel (a)], which inherit the sharp edges owing to the conformal character of the transformations. The final structure is thus a surface where SPs can be excited, and which are localized at the surface and decay evanescently away into the metal and dielectric half-spaces. Hence, it may seem at first sight that these modes are two-dimensional, that is, characterized by two wave-vectors parallel to the metal surface. However, due to the conformal character of the transformations, the singular metasurface inherits the spectral properties of the slab array, which supports three-dimensional modes, characterized by the two wave-vectors in the plane where the slabs extend to infinity and by a third one along the direction where they are periodic. As a consequence, the modes supported by the singular surface are also characterized by three wave-vectors, with the third one being inherited from the transformed structure and associated to an extra dimension that is compacted into the singularities \cite{Pendry2017}. This has remarkable consequences in the optical spectrum of the singular metasurface: the extra wave-vector is not subject to a selection rule, and therefore there is a mode available at every frequency, which results in a broadband absorption spectrum. In other words, these gratings are black (or gray) while conventional gratings, which have discrete absorption lines, are coloured.  

The broadband spectral response of a singular silver metasurface of period $10$ nm is shown in Fig. \ref{fig:singulargratings}(a), where the normal incidence reflectivity spectrum is plotted as a solid grey line. A continuous band of low reflectivity (high absorption) can be seen, which corresponds to the excitation of the antisymmetric mode between a cut-off frequency and the SP frequency ($\omega_{sp} = \omega_p/\sqrt{2}$). This is in striking difference with the results discussed in Section \ref{sec:gratings} for a non-singular plasmonic grating, which features a discrete series of resonant modes [see Fig. \ref{fig:gratings} (c)]. The optical response of the singular structure was calculated analytically by representing a plane wave incident on the metasurface as an array of magnetic line currents located at infinity \cite{Yang2018}. These sources are mapped to a periodic array of sources in the slab array frame, where the power flow carried by the excited SPs as they travel towards $\pm\infty$ is calculated. Next, the singular periodic surface is represented by a flat surface with an effective conductivity, which can be determined through conservation of energy, and from which the reflectivity of the singular metasurface is calculated. We note that numerical calculations of this system are not possible due to the singular character of the geometry. 

In practice, perfect singularities are not possible to realize. Even if recent advances in nanofabrication enable the experimental realization of plasmonic structures with high precision \cite{Benz2016picocavities}, achieving a perfectly singular point will always be limited by the discrete nature of the electron gas, which prevents the existence of a perfect singularity where the electron density would diverge. The finite screening length of metals ($\sim 0.1 $ nm in noble metals) limits the size where electrons can accumulate in the singularity, and prevents the density from blowing up. These \textit{non-local effects} effectively blunt the singularities which has a strong impact on the optical response of singular metasurfaces \cite{YangPhysRevB2019}. This can be seen in Fig. \ref{fig:gratings} (a), which presents the reflectivity for the singular metasurface using a nonlocal dielectric permittivity as solid blue lines, with red dashed lines obtained from numerical simulations also shown for comparison. The hydrodynamic model was assumed, and the value of $\beta$, a parameter that determines the screening length and hence the extent of the singularity, was tuned down artificially from a realistic value for silver (left panel) to a very low value (right panel). Nonlocality blunts the singularities, which map to slabs of finite length in the transformed frame. These are cavities for the SPs, which discretizes the spectrum and a set of reflectivity dips are observed (left). As the local regime is approached, the singularity is effectively sharper and in the transformed frame the cavities are longer, such that the structure supports more and more resonances (middle), tending towards the continuum obtained the local approximation when nonlocality is very small (right). The remarkable influence of nonlocality in the optical spectrum of the singular metasurface indicates that they could be used as a platform to probe nonlocality in metals.  

A second instance of singular metasurfaces that can be smoothly approached can be realized in graphene as proposed in Ref. \cite{Pendry2017}. In this case, the conformal transformation introduced in Section \ref{sec:gratings} was adapted to generate a surface with singularities in the form of touching points rather than sharp edges. This is done by first renormalising the whole structure through the introduction of a new length scale in the slab frame,  $d$ (the period of slab that maps into the length between two branch points in the transformed geometry). With this, the transformation reads as,
\begin{equation}
    z = \frac{d'}{2\pi} \ln \left(\frac{1}{e^{2\pi w/d}-iw_0} +iy_0 \right),
\end{equation}
with $y_0$ now defined as $y_0 = w_0/( \exp[4\pi(u_0 + \delta)/d] − w^2_0)$. Then the origin of the inversion is taken at a point very close to one of the surfaces ($w_0\rightarrow 1$), which generates a grating with vanishing thickness at the valley points [see inset of Fig. \ref{fig:singulargratings}(b)]. Similar to the non-singular grating, the free parameter in the transformation, $w_0$, determines the shape of the grating, and the singular behaviour, with $w_0\rightarrow 1$ representing the singularity where the two surfaces touch, or where the doping approaches zero in the case of graphene. 

Figure \ref{fig:singulargratings}(b) presents the absorption spectrum of singular metasurfaces realized on graphene. The singularity is approached by keeping the same maximum conductivity value  while reducing the minimum value, which is suppressed from the left to the right panels. When the grating is far from singular, the spectrum shows a discrete set of peaks corresponding to plasmonic resonances of increasing order (left). As the singularity is approached, more and more resonances appear in the spectrum (medium), and when the system is very close to being singular, the spectrum tends to a broadband of continuous absorption. These results assume a realistic value of the loss (mobility $m=10^4$ cm$^2$/(V$\cdot$s)), and we stress that increasing the loss ($m=3\times10^3$ cm$^2$/(V$\cdot$s)) further merges the peaks into the broadband. This sytsem has been suggested as a tunable ultra-thin broadband absorber for THz waves \cite{galiffi2018broadband}. Similar to the singular silver surface, the broadband absorption spectrum can be explained by means of an extra dimension compacted in the singularity. This additional dimension is inherited from the periodicity introduced in the slab frame ($d$), which tends to infinity, while the dimension of the slab along its length  is itself infinite. This results in a hidden dimension in the singularity in the grating frame, where incident radiation can satisfy the dispersion relation over a continuous frequency band. In fact, as the period in the slab frame increases, the modes are discretized in a smaller Brillouin zone. As a consequence, SP modes at larger wavevectors are available at lower and lower frequencies. The large confinement characteristic of these modes is responsible for the large absorptions seen in the singular metasurfaces \cite{Galiffi2019epj}. Finally, we remark that these singular graphene metasurfaces provide a platform for the study of nonlocality in graphene, which is stronger when the doping is lower. The SPs propagating towards the singularity are a sensitive probe of nonlocal effects in graphene, which would become observable in far field measurements.   

In this Section we have reviewed the use of TO to design plasmonic metasurfaces, and the proposal of singular plasmonic metasurfaces which hide an extra dimension in the singularity and which be used as ultrathin broadband absorbers. In the following, we turn our attention into a different area of nanophotonics, that of exctiton-plasmon interactions in nanocavities.

\section{Exciton-plasmon strong coupling}

In recent years, much theoretical efforts have focused on developing a general methodology for
the expansion of the Dyadic Green's functions in open, lossy and
dispersive systems in terms of a discrete set of EM modes.
However, although this is currently a topic of intense activity,
there is not yet a consensus about the precise definition of these
EM modes, their associated eigenfunctions and eigenvalues. As a
consequence, various terms, such as resonant states~\cite{muljarov2016resonant}, generalized
normal~\cite{chen2019generalizing} or
quasinormal~\cite{sauvan2013theory,kristensen2013modes} modes, have been coined lately to refer to them. Indeed, the
conception of a theoretical framework allowing for a general Green's
function decomposition would mean a significant advance in
multiple areas. The investigation of quantum optical phenomena in
plasmonic~\cite{yang2015analytical,hughes2018quantized} and
metallodielectric~\cite{franke2019quantization} nanocavities is among them. It
would allow for a convenient quantization of subwavelength EM
fields avoiding the enormous number of degrees of freedom inherent
to macroscopic quantum electrodynamics
calculations ~\cite{dung1998three}.

As discussed in Section \ref{sec:Intro}, TO has been used in the past to obtain analytical descriptions of the light collection and concentration by a wide range of nanoparticle geometries. In this section, we
discuss the application of similar methods to build 3D~\cite{li2016transformation,li2018symmetry} (subsection~\ref{sec:SC1}) and 2D~\cite{pacheco2016description,pacheco2017aluminum,cuartero2018light,cuartero2019dipolar} (subsection~\ref{sec:SC2}) models of the
response of similar structures to point-like EM sources, such as quantum emitters (QEs), placed in their vicinity. This way, TO
provides analytical insights into the Dyadic Green's function for
these systems. Importantly, this approach also reveals its
convenient decomposition and the proper definition of modal
eigenvalues and eigenfunctions. Specifically, TO has been employed in the investigation of plasmon-exciton
interactions in nanocavities, accounting for the full richness of
the EM spectrum in these devices and revealing the
conditions yielding strong coupling at the single QE level. Note that, contrary to nanoantennas, where the objective is
enhancing the near- to far-field transfer of EM energy, this must be reduced in
nanocavities for strong light-matter coupling. This means that the quasistatic approximation is an optimum
starting point for the analysis of these phenomena.

\begin{figure}[!t]
\includegraphics[width=0.9\linewidth]{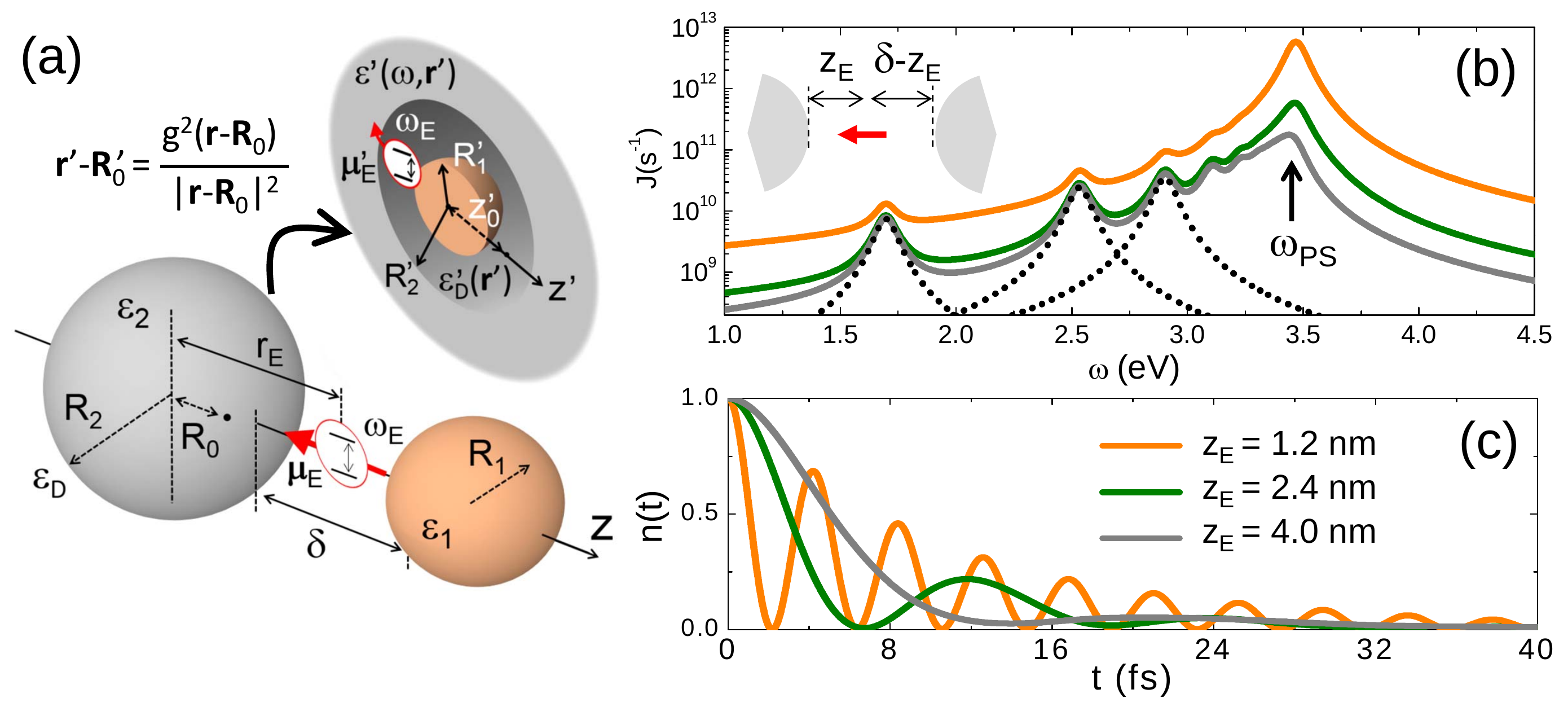}
\vspace{-0.35 cm} \caption{(a) Left: Sketch of a single QE with
transition dipole moment $\boldsymbol{\mu}_{\rm E}$ and frequency $\omega_{\rm E}$ placed within the
gap between two nanospheres. Right: Spherically symmetric geometry
obtained from the inversion of the original system. (b)
Quasistatic spectral density at three different QE positions, see
panel (c), along the $z$-axis and within the 8 nm gap between two
Ag spheres ($R=120$ nm). (c) Population dynamics, $n(t)$
($n(0)=1$), for a QE at resonance with the plasmonic pseudomode of
the gap nanocavity, $\omega_{\rm E}=\omega_{\rm PS}$, for three
$z_{\rm E}$ values. Adapted with permission~\cite{li2016transformation,li2018symmetry}.}
\label{fig:SC1}
\end{figure}

\subsection{Three-dimensional model}\label{sec:SC1}
We consider first a nanocavity composed by two identical metallic
spheres of radius $R$, with a Drude-like permittivity, separated by a nanometric gap, $\delta$. As illustrated in
Fig.~\ref{fig:SC1}(a), they can be transformed into a metal-dielectric-metal spherically-symmetric geometry under an inversion. This requires choosing judiciously the inversion point, $R_0$~\cite{li2016transformation}. As a result of the mapping, the scalar permittivity in the transformed frame acquires a spatial dependence of the form $\epsilon'(\omega,{\bf r}')=g^2\epsilon(\omega,{\bf r}({\bf
r}'))/|{\bf r}'-{\bf R}_0'|^2$, where $\epsilon(\omega,{\bf r})$ is the original dielectric constant distribution. Note that the EM fields do not depend on the choice of $g$, a constant setting the length-scale of the mapping. It can be
proven~\cite{fernandez2010collection} that the quasistatic
potential in the inverted system can be written as
$\Phi'(\mathbf{r}')=
|\mathbf{r}'-\mathbf{R}'_0|\phi'(\mathbf{r}')$ where
$\phi'(\mathbf{r}')$ is a solution of Laplace's Equation in the
primed frame. The potential in the transformed geometry can then be obtained by expanding $\Phi'(\mathbf{r}')$ in terms of spherical solutions of Laplace's Equation (labelled with degree $l$ and order $m$ of spherical harmonics) and imposing continuity conditions at the concentric spherical boundaries. Once this is known, the potential in the original frame is given by $\Phi(\mathbf{r})=\Phi'(\mathbf{r}'(\mathbf{r}))$. 

Using the TO approach briefly introduced above, the Dyadic Green's function for the system can be calculated by introducing a point-like dipole source, $\Phi_{\rm E}(\mathbf{r})$,  modelling a QE as the EM excitation in the original frame~\cite{li2018symmetry}. By imposing $m=0$ in the potential expansion, the calculations simplify significantly. With this restriction, only  sources located along the $z$-direction (the nanocavity axis) and oriented parallel to it can be treated. Note that the interaction with the SPs supported by the spheres is maximum in these conditions.  The component of the scattering Dyadic Green's function governing the QE-SP interaction is $G^{\rm sc}_{zz}(\omega,\mathbf{z}_{\rm E},\mathbf{z}_{\rm E})=\frac{\epsilon_0}{\mu_{\rm E}}(\frac{c}{\omega})^{2}|\partial_z\Phi_{\rm sc}(\omega,\mathbf{z}_{\rm E})|$, where $\mu_{\rm E}$ is the QE dipole moment, $z_{\rm E}$ its position, and $\Phi_{\rm sc}(\omega,\mathbf{r})=\Phi(\omega,\mathbf{r})-\Phi_{\rm E}(\mathbf{r})$. Note that, for clarity, the spectral dependence of the scattered quasistatic potential is indicated, which originates from the presence of $\epsilon(\omega,\mathbf{r})$ in the continuity equations.  

The spectral density~\cite{breuer2002theory}, the physical magnitude that weights light-matter coupling in the nanocavity, can be expressed as~\cite{gonzalez2014reversible,delga2014quantum} 
\begin{equation}
J(\omega)=\frac{\gamma_{\rm E}(\omega)}{2\pi}P(\omega)=\frac{\mu_{\rm
E}^2\omega^2}{\pi\epsilon_0\hbar c^2} {\rm Im}\{G^{\rm
sc}_{zz}(\omega,\mathbf{r}_{\rm E},\mathbf{r}_{\rm
E})\}=\sum_{l=1}^\infty\sum_{\sigma=\pm
1}\frac{g_{l,\sigma}^2}{\pi}\frac{\gamma_{\rm D}/2}{(\omega-\omega_{l,\sigma})^2+(\gamma_{\rm D}/2)^2}\label{eq:J}
\end{equation}
where $\gamma_{\rm E}(\omega)=\omega^3\mu_{\rm E}^2/3\pi\epsilon_0\hbar c^3$ is the spontaneous decay rate of the QE ($\omega_{\rm E}=\omega$) in free space and $P(\omega)$ the Purcell enhancement induced by the nanocavity~\cite{novotny2012principles,giannini2011plasmonic}. Note that $J(\omega)$  is, except for a factor, the QE decay rate in the plasmonic environment. The right-hand side in Eq.~\ref{eq:J} results from the Green's function decomposition given by the TO approach. In the limit of small gap sizes, $\rho=\delta/R<<1$, the SPs can be labelled in terms of their angular momentum $l$, and their even/odd parity across the gap, $\sigma$~\cite{li2016transformation}. This way, analytical expressions for the SP frequencies, $\omega_{l,\sigma}$, and SP-QE coupling constants, $g_{l,\sigma}$, are obtained. Note that $\gamma_{\rm D}$ in Eq.~\ref{eq:J} is the absorption rate in the metal Drude permittivity, the only damping mechanism in the quasistatic regime. 

Figure~\ref{fig:SC1}(b) plots $J(\omega)$  at the 8 nm gap between two
Ag spheres of radius $120$ nm. Three different QE positions are considered, $z_{\rm E}$: 4 nm (the gap center, in grey), 2.4 nm (green) and 1.2 nm (orange). The QE dipole moment is set to $\boldsymbol{\mu}_{\rm E}=1.5\,{\rm
e}\cdot{\rm nm}$. The first three even ($\sigma=1$) Lorentzian terms in the expansion in Eq.~\ref{eq:J} are plotted in blue dashed lines (they are the same for all $z_{\rm E}$). These correspond to the lowest energy, most radiative SP modes which govern the absorption properties of the sphere dimer under plane wave illumination~\cite{pendry2013capturing,li2018symmetry}. The spectral density presents a much stronger feature at higher frequencies, this is the plasmonic pseudomode, which emerges as a result of the spectral overlapping (within a frequency window $\gamma_{\rm D}$) of SP modes with high angular momentum (large $l$)~\cite{delga2014quantum}. Note that $\omega_{\rm PS}$ lies in the vicinity of the quasistatic SP frequency for the metal permittivity. Fig.~\ref{fig:SC1}(b) shows that $J(\omega_{\rm PS})$ increases as the QE is displaced away from the gap center and approaches one of the sphere surfaces (it couples more efficiently to SPs with shorter evanescent tails into the gap region), while the contribution due to low-frequency SPs do not vary with $z_{\rm E}$. 

Figure~\ref{fig:SC1}(c) renders the QE exciton population as a function of time, $n(t)$ in a spontaneous emission configuration ($n(0)=1$) for the three positions in panel (b) and for $\omega_{\rm E}=\omega_{\rm PS}$. The popuplation dynamics are calculated using the Wigner-Weisskopf Equation~\cite{breuer2002theory} fed with the TO-calculated spectral densities. The QE-SP interaction is in the weak-coupling regime at the gap center (grey) and $n(t)$ decays monotonically. However, for QEs away from the gap center, Rabi oscillations emerge in $n(t)$, and become stronger with smaller $z_{\rm E}$. These are the fingerprint of the onset of strong coupling, and reveal that the population is transferred back and forth between the QE and the nanocavity (the pseudomode it supports) several times before its decay due to metal absorption. Fig.~\ref{fig:SC1}(c) demonstrates that plasmon-exciton polaritons at the single QE level can be formed in nanocavities with large (4 nm) gaps by displacing the emitter position away from the gap center.   

\subsection{Two-dimensional model}\label{sec:SC2}

The 3D model in the previous section presents several limitations. It yields analytical expressions only for dipolar sources located along, and oriented parallel to, the symmetry axis of cavities with small $\rho=\delta/R$. Moreover, the description of microscopic sources of higher order than dipolar ones cannot be handled analytically either. Finally, it is purely quasi-static and therefore does not provide any insight into far-field magnitudes, which are instrumental for the experimental probing of hybrid QE-SP systems. In the following, we show how these constraints can be overcome by considering a 2D model of the nanocavity, in which translational invariance along $y$-direction of the EM fields is assumed. Importantly, this approximation is justified by the remarkable similarity between plasmon-exciton strong-coupling phenomenology in 2D and 3D geometries~\cite{demetriadou2017spatiotemporal}.

\begin{figure}[!t]
\hspace{1 cm}
\includegraphics[width=0.9\linewidth]{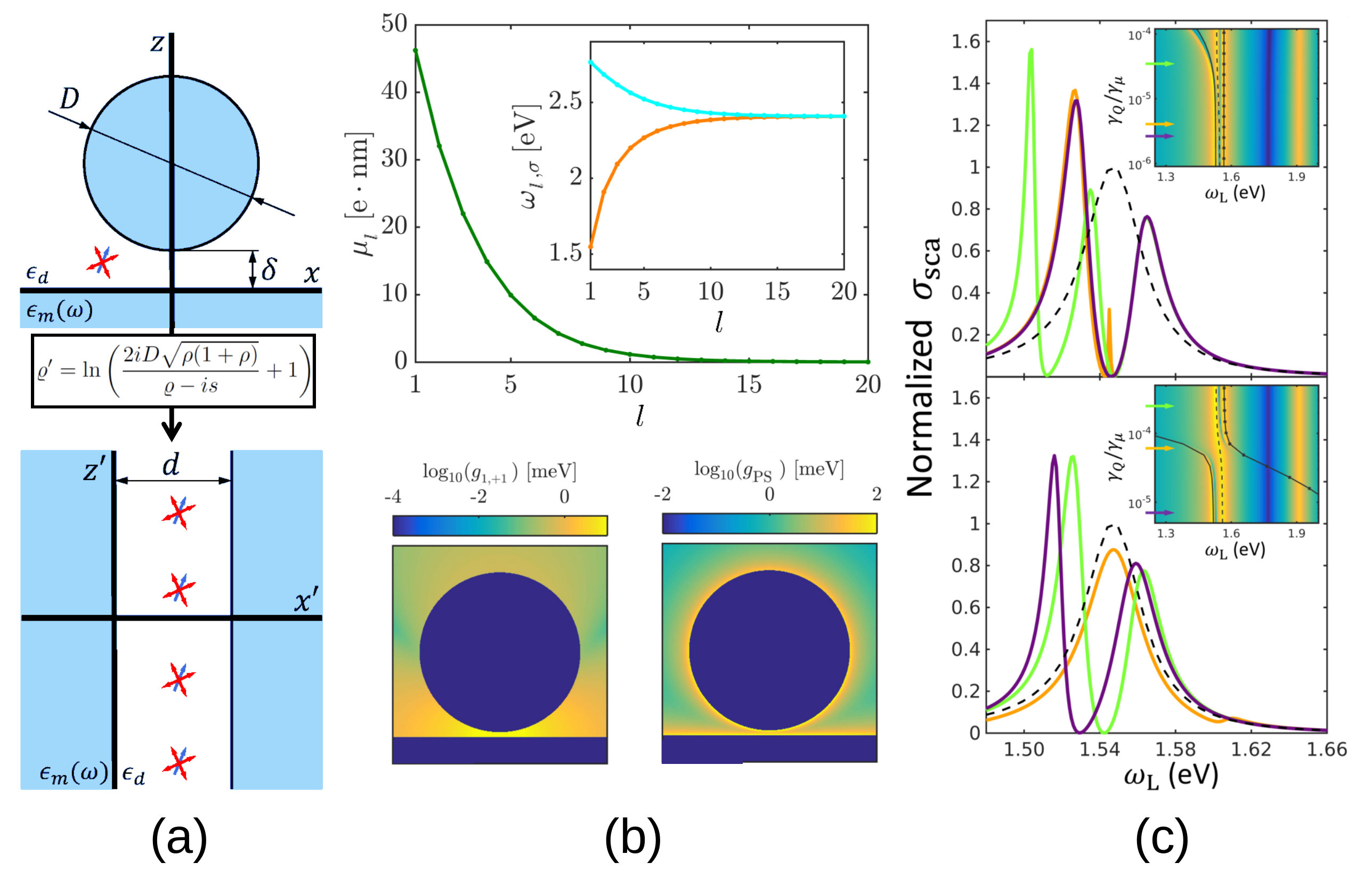}
\vspace{-0.2 cm}\caption{(a) 2D mapping between a Ag NPoM geometry and a silver-dielectric-silver waveguide. (b) Top: Even SP dipole moment versus index $l$. Inset: Even (orange) and odd (blue) SP frequencies ($D=30$ nm, $\delta=0.9$ nm). Bottom: Coupling strength maps for vertically oriented QES and the lower-order dipolar SP mode (left) and the plasmonic pseudomode (right). (c) Scattering spectra around $\omega_{1+1}=\omega_\mu$ for three-level QEs (sustaining one dipolar and one quadrupolar transition) with $z_E=\delta/2$ and $\omega_Q=\omega_{1,+1}$ (top) and $z_E=7\delta/8$ and $\omega_Q=\omega_{\rm PS}$. Adapted with permission~\cite{cuartero2018light,cuartero2019dipolar}.}
\label{fig:SC2}
\end{figure}

Figure~\ref{fig:SC2}(a) shows how the 2D version of a nanoparticle-on-a-mirror (NPoM) geometry can be transformed into a metal-dielectric-metal waveguide under a logarithmic conformal map ($\varrho^{(\prime)}=x^{(\prime)}+iz^{(\prime)}$) with $D=2R$ and $s=\delta+D\sqrt{\rho}/(\sqrt{2+\rho}+\sqrt{\rho})$)~\cite{aubry2011plasmonic}. The original EM point-like source transforms into an array of coherent identical sources, which makes the transformed system periodic. This periodicity provides again with appropriate indices for the SP modes: the Bloch band index, $l$, and, similarly to the 3D case, the parity with respect to the waveguide symmetry plane, $\sigma$~\cite{cuartero2018light}. The spectral densities can be calculated from the 2D model by using the first equality in Eq.~\ref{eq:J}, fed with 2D calculations of the Purcell enhancement $P(\omega)=\frac{8\epsilon_0}{\mu_{\rm E}^2}\Big(\frac{c}{\omega}\Big)^2 {\rm
Im}\{\boldsymbol{\mu}_{\rm E} \nabla\ \Phi(\textbf{
r},\omega)|_{\textbf{r}_{E}}\}$, where $\textbf{r}=(x,z)$ and $\textbf{ r}_{E}$ is the position of
the emitter in the $xz$-plane. This simplified model makes it possible treating quadrupolar exciton transitions in QEs in an analytical fashion as well~\cite{cuartero2019dipolar}. Once 2D Purcell factors are known, they are combined with 3D free-space decay rates in Eq.~\ref{eq:J}.

Implementing radiation reaction corrections in the 2D model~\cite{aubry2010conformal}, the radiative decay rate for the even ($\sigma=+1$) SPs supported by NPoM cavities, $\gamma^{\rm r}_{l,+1}$, can be calculated~\cite{demetriadou2017spatiotemporal} (note that, by symmetry, $\gamma^{\rm r}_{l,-1}=0$). With this theory refinement, the spectral width of the Lorentzian terms in Eq.~\ref{eq:J} acquire the form $\gamma_{l,\sigma}=\gamma_{\rm D}+\gamma^{\rm r}_{l,+1}\delta_{\sigma,+1}$. Moreover, using the method of images, the dipolar moment of even SPs can be extracted out of $\gamma^{\rm r}_{l,+1}$~\cite{cuartero2019dipolar}. The top panel of Figure~\ref{fig:SC2} renders the SP dipolar moment versus index $l$ for the NPoM cavity in panel (a). As expected, they decrease as the mode order increases, and the SPs contributing to the plasmonic pseudomode are completely dark. The inset plots the SP frequencies for even (orange) and odd (blue) parity, showing that both branches overlap for large $l$. To illustrate the power of the TO approach, the bottom panels in Figure~\ref{fig:SC2}(b) display the coupling strength maps for vertically oriented QEs and for the lowest even SP mode ($g_{1,+1}$, left) and the plasmonic pseudomode ($g_{\rm PS}$, right). The former is focused at the gap region, where it becomes rather uniform, and decays away from it. The latter is tightly confined to the metal boundaries, both at the particle and flat substrate, and presents little sensitivity to plasmon hybridization effects across the gap of the cavity.

Using the TO-calculated SP frequencies, $\omega_{l,\sigma}$, and the QE-SP coupling constants, $g_{l,\sigma}$, the Jaynes-Cummings Hamiltonian~\cite{breuer2002theory} describing plasmon-exciton interactions in the NPoM cavity can be parametrized
\begin{equation}
\hat{H}=\omega_{\rm E}\hat{\sigma}^{\dag}\hat{\sigma}+\sum_{l,\sigma}\omega_{l,\sigma}\hat{a}^{\dag}_{l,\sigma}\hat{a}_{l,\sigma}+\sum_{l,\sigma}g_{l,\sigma}[\hat{\sigma}^{\dag}\hat{a}_{l,\sigma}+\hat{\sigma}\hat{a}_{l,\sigma}^{\dag}],\label{eq:H}
\end{equation}
where $\hat{\sigma}$ and $\hat{a}_{l,\sigma}$ are the QE and SP annihilation operators. Eq.~\ref{eq:H} illustrates the exploitation of TO as a tool for the quantization of the complex plasmonic spectrum supported by the NPoM cavity. Moreover, through Lindblad terms weighted by the SP damping rates, $\gamma_{l,\sigma}$, we can account for plasmonic losses in a master equation description of the system~\cite{cuartero2019dipolar}.   

Figure~\ref{fig:SC2}(c) shows scattering spectra for the cavity in panel (a) coupled to a QE modelled as a three-level system sustaining two, one dipolar and one quadrupolar, exciton transitions. Note that the latter is dark and could not be accessed by propagating EM fields. By adding a coherent driving term~\cite{esteban2014strong} to Eq.~\ref{eq:H}, we can describe the illumination of the system by a laser field of frequency $\omega_L$ around the lowest SP resonance, $\omega_{1,+1}$. The far-field scattering spectrum can be computed as the square of the steady-state expectation value of the total dipole moment operator, $\hat{M}=\sum_l\mu_{l}\hat{a}_{l+1}+\mu_{\rm E}\hat{\sigma}_{\mu}$~\cite{saez2017enhancing}. The bare nanocavity is shown in dashed black lines. The QE is located at the center of the gap in the top panel and displaced along $z$-directon in the bottom one. In both cases, the dipolar transition is set at resonance with $\omega_{1,+1}$. In absence of the quadrupolar transition, a Rabi doublet~\cite{laussy2008strong} is apparent in the spectra of the hybrid system, the fingerprint of the onset of QE-SP strong coupling (see violet lines)~\cite{saez2018photon}. By increasing the quadrupole moment of the dark QE transition (orange and green lines), this spectral profile is modified in opposite ways. In the top panel, a third peak emerges at the dip between the Rabi maxima, whereas a single peak (resembling the bare cavity) is observed in the bottom panel. These spectra show how TO allows exploring the remarkable effect that dark excitons can have in QE-SP interactions in the strong-coupling regime~\cite{cuartero2018light}.   

\section{Conclusions}
In this review we have discussed the theory of transformation optics and its applications. We have first reviewed its impact in metamaterial science for the design of electromagnetic devices and other aspects, such as the control of surface waves or its extension into other realms of wave physics. Next, we have revised how transformation optics has provided a set of analytical techniques for investigating complex problems in plasmonics. We have then focused on the application of this theoretical framework to the analytical treatment of two open problems of much relevance in current theoretical nanophotonics. 

On the one hand, we have shown how transformation optics allows for the design of plasmonic metasurfaces with predictable optical responses inherited from a transformed structure with more symmetries. We have also reviewed the proposal of singular plasmonic metasurfaces in the form of subwavelength metal gratings with sharp edges or graphene metasurfaces with points of vanishing doping level. These singular structures are a realization of compacted dimensions, provide macroscopic signatures of nonlocality, and could be used as ultrathin broadband absorbers. 

On the other hand, we have presented the description of plasmon-exciton coupling in nanocavities by means of transformation optics calculations. We have discussed the insights that this tool provides into both near- and far-field physical magnitudes, such as the exciton dynamics and the scattering spectrum. Finally, we have shown that this tool enables the analytical parametrization of the Jaynes-Cummings Hamiltonian describing light-matter interactions in these hybrid nanometric systems.    

\section*{Acknowledgements}
P.A.H. acknowledges funding from Funda\c c\~ao para a Ci\^encia e a Tecnologia and Instituto de Telecomunica\c c\~oes under project CEECIND/03866/2017, and The Leverhulme Trust. A. I. F.-D. acknowledges funding from the Spanish MINECO and MICINN
under Contracts FIS2015-64951-R and RTI2018-099737-B-I00, as well as the “Mar\'{i}a de Maeztu” programme for Units of Excellence in R\&D (MDM-2014-0377).

\bibliography{review_TO_comptesrend}

\begin{thebibliography}{100}

\bibitem{Ward1996}
A~J Ward and John~B. Pendry.
\newblock {Refraction and Geometry in Maxwell's Equations}.
\newblock {\em J. Mod. Opt.}, 43(4):773 ---- 793, 1996.

\bibitem{Pendry2006}
John~B. Pendry, D~Schurig, and D~R Smith.
\newblock {Controlling Electromagnetic Fields}.
\newblock {\em Science (New York, N.Y.)}, 312(5781):1780--2, jun 2006.

\bibitem{Chen2010}
Huanyang Chen, C~T Chan, and Ping Sheng.
\newblock {Transformation optics and metamaterials.}
\newblock {\em Nat. Mater.}, 9(5):387--96, may 2010.

\bibitem{Huidobro2010}
Paloma~A. Huidobro, Maxim~L Nesterov, Luis Mart{\'{i}}n-Moreno, and
  Francisco~J. Garc{\'{i}}a-Vidal.
\newblock {Transformation optics for plasmonics.}
\newblock {\em Nano Lett.}, 10(6):1985--90, jul 2010.

\bibitem{sun2017transformation}
Fei Sun, Bin Zheng, Hongsheng Chen, Wei Jiang, Shuwei Guo, Yichao Liu, Yungui
  Ma, and Sailing He.
\newblock Transformation optics: From classic theory and applications to its
  new branches.
\newblock {\em Laser \& Photonics Reviews}, 11(6), 11 2017.

\bibitem{mccall2010spacetime}
Martin~W McCall, Alberto Favaro, Paul Kinsler, and Allan Boardman.
\newblock A spacetime cloak, or a history editor.
\newblock {\em Journal of optics}, 13(2):024003, 2010.

\bibitem{fridman2012demonstration}
Moti Fridman, Alessandro Farsi, Yoshitomo Okawachi, and Alexander~L Gaeta.
\newblock Demonstration of temporal cloaking.
\newblock {\em Nature}, 481(7379):62, 2012.

\bibitem{leonhardt2006general}
Ulf Leonhardt and Thomas~G Philbin.
\newblock General relativity in electrical engineering.
\newblock {\em New Journal of Physics}, 8(10):247, 2006.

\bibitem{greenleaf2007electromagnetic}
Allan Greenleaf, Yaroslav Kurylev, Matti Lassas, and Gunther Uhlmann.
\newblock Electromagnetic wormholes and virtual magnetic monopoles from
  metamaterials.
\newblock {\em Physical Review Letters}, 99(18):183901, 2007.

\bibitem{smolyaninov2010metric}
Igor~I Smolyaninov and Evgenii~E Narimanov.
\newblock Metric signature transitions in optical metamaterials.
\newblock {\em Physical Review Letters}, 105(6):067402, 2010.

\bibitem{thompson2010completely}
Robert~T Thompson, Steven~A Cummer, and Joerg Frauendiener.
\newblock A completely covariant approach to transformation optics.
\newblock {\em Journal of Optics}, 13(2):024008, 2010.

\bibitem{horsley2016wave}
SAR Horsley, CG~King, and TG~Philbin.
\newblock Wave propagation in complex coordinates.
\newblock {\em Journal of Optics}, 18(4):044016, 2016.

\bibitem{horsley2015spatial}
SAR Horsley, M~Artoni, and GC~La~Rocca.
\newblock Spatial kramers--kronig relations and the reflection of waves.
\newblock {\em Nature Photonics}, 9(7):436, 2015.

\bibitem{ruter2010observation}
Christian~E R{\"u}ter, Konstantinos~G Makris, Ramy El-Ganainy, Demetrios~N
  Christodoulides, Mordechai Segev, and Detlef Kip.
\newblock Observation of parity--time symmetry in optics.
\newblock {\em Nature physics}, 6(3):192, 2010.

\bibitem{castaldi2013p}
Giuseppe Castaldi, Silvio Savoia, Vincenzo Galdi, Andrea Alu, and Nader
  Engheta.
\newblock P t metamaterials via complex-coordinate transformation optics.
\newblock {\em Physical review letters}, 110(17):173901, 2013.

\bibitem{mitchell2013perfect}
RC~Mitchell-Thomas, TM~McManus, O~Quevedo-Teruel, SAR Horsley, and Y~Hao.
\newblock Perfect surface wave cloaks.
\newblock {\em Physical review letters}, 111(21):213901, 2013.

\bibitem{Viaene2016}
Sophie Viaene, Vincent Ginis, Jan Danckaert, and Philippe Tassin.
\newblock Transforming two-dimensional guided light using nonmagnetic
  metamaterial waveguides.
\newblock {\em Phys. Rev. B}, 93:085429, Feb 2016.

\bibitem{quevedo2013transformation}
Oscar Quevedo-Teruel, Wenxuan Tang, Rhiannon~C Mitchell-Thomas, Amy Dyke, Hazel
  Dyke, Lianhong Zhang, Sajad Haq, and Yang Hao.
\newblock Transformation optics for antennas: why limit the bandwidth with
  metamaterials?
\newblock {\em Scientific reports}, 3:1903, 2013.

\bibitem{Schurig2006}
D~Schurig, J~J Mock, B~J Justice, Steven~A. Cummer, John~B. Pendry, A~F Starr,
  and David~R. Smith.
\newblock {Metamaterial Electromagnetic Cloak at Microwave Frequencies}.
\newblock {\em Science}, 314:977, 2006.

\bibitem{cummer2007one}
Steven~A Cummer and David Schurig.
\newblock One path to acoustic cloaking.
\newblock {\em New Journal of Physics}, 9(3):45, 2007.

\bibitem{chen2007acoustic}
Huanyang Chen and CT~Chan.
\newblock Acoustic cloaking in three dimensions using acoustic metamaterials.
\newblock {\em Applied physics letters}, 91(18):183518, 2007.

\bibitem{Zigoneanu2014}
Lucian Zigoneanu, Bogdan-Ioan Popa, and Steven~A. Cummer.
\newblock Three-dimensional broadband omnidirectional acoustic ground cloak.
\newblock {\em Nature Materials}, 13:352 EP --, Mar 2014.

\bibitem{zhang2008cloaking}
Shuang Zhang, Dentcho~A Genov, Cheng Sun, and Xiang Zhang.
\newblock Cloaking of matter waves.
\newblock {\em Physical Review Letters}, 100(12):123002, 2008.

\bibitem{guenneau2012transformation}
Sebastien Guenneau, Claude Amra, and Denis Veynante.
\newblock Transformation thermodynamics: cloaking and concentrating heat flux.
\newblock {\em Optics Express}, 20(7):8207--8218, 2012.

\bibitem{Milton2006}
Graeme~W Milton, Marc Briane, and John~R Willis.
\newblock On cloaking for elasticity and physical equations with a
  transformation invariant form.
\newblock {\em New Journal of Physics}, 8(10):248--248, oct 2006.

\bibitem{stenger2012experiments}
Nicolas Stenger, Manfred Wilhelm, and Martin Wegener.
\newblock Experiments on elastic cloaking in thin plates.
\newblock {\em Physical Review Letters}, 108(1):014301, 2012.

\bibitem{buckmann2014elasto}
T~B{\"u}ckmann, M~Thiel, M~Kadic, R~Schittny, and M~Wegener.
\newblock An elasto-mechanical unfeelability cloak made of pentamode
  metamaterials.
\newblock {\em Nature communications}, 5:4130, 2014.

\bibitem{buckmann2015mechanical}
Tiemo B{\"u}ckmann, Muamer Kadic, Robert Schittny, and Martin Wegener.
\newblock Mechanical cloak design by direct lattice transformation.
\newblock {\em Proceedings of the National Academy of Sciences},
  112(16):4930--4934, 2015.

\bibitem{Brun2009}
M.~Brun, S.~Guenneau, and A.~B. Movchan.
\newblock Achieving control of in-plane elastic waves.
\newblock {\em Applied Physics Letters}, 94(6):061903, 2009.

\bibitem{Guenneau2014}
S.~Br\^ul\'e, E.~H. Javelaud, S.~Enoch, and S.~Guenneau.
\newblock Experiments on seismic metamaterials: Molding surface waves.
\newblock {\em Phys. Rev. Lett.}, 112:133901, Mar 2014.

\bibitem{liu2010transformational}
Yongmin Liu, Thomas Zentgraf, Guy Bartal, and Xiang Zhang.
\newblock Transformational plasmon optics.
\newblock {\em Nano letters}, 10(6):1991--1997, 2010.

\bibitem{Kadic2012}
Muamer Kadic, S{\'{e}}bastien Guenneau, Stefan Enoch, Paloma~A. Huidobro, Luis
  Mart{\'{i}}n-Moreno, Francisco~J. Garc{\'{i}}a-Vidal, Jan Renger, and Romain
  Quidant.
\newblock {Transformation plasmonics}.
\newblock {\em Nanophotonics}, 1(1):51--64, jan 2012.

\bibitem{Huidobro2011}
Paloma~A. Huidobro, M~L Nesterov, Luis Mart{\'{i}}n-Moreno, and Francisco~J.
  Garc{\'{i}}a-Vidal.
\newblock {Moulding the flow of surface plasmons using conformal and
  quasiconformal mappings}.
\newblock {\em New J. Phys.}, 13(3):033011, mar 2011.

\bibitem{Renger2010}
Jan Renger, Muamer Kadic, Guillaume Dupont, Srdjan~S. A\'{c}imovi\'{c},
  S\'{e}bastien Guenneau, Romain Quidant, and Stefan Enoch.
\newblock Hidden progress: broadband plasmonic invisibility.
\newblock {\em Opt. Express}, 18(15):15757--15768, Jul 2010.

\bibitem{zentgraf2011plasmonic}
Thomas Zentgraf, Yongmin Liu, Maiken~H Mikkelsen, Jason Valentine, and Xiang
  Zhang.
\newblock Plasmonic luneburg and eaton lenses.
\newblock {\em Nature nanotechnology}, 6(3):151, 2011.

\bibitem{novotny2012principles}
Lukas Novotny and Bert Hecht.
\newblock {\em Principles of nano-optics}.
\newblock Cambridge university press, 2012.

\bibitem{aubry2010conformal}
Alexandre Aubry, Dang~Yuan Lei, Stefan~A Maier, and JB~Pendry.
\newblock Conformal transformation applied to plasmonics beyond the quasistatic
  limit.
\newblock {\em Physical Review B}, 82(20):205109, 2010.

\bibitem{pendry2012transformation}
JB~Pendry, A~Aubry, DR~Smith, and SA~Maier.
\newblock Transformation optics and subwavelength control of light.
\newblock {\em science}, 337(6094):549--552, 2012.

\bibitem{luo2013harvesting}
Yu~Luo, Rongkuo Zhao, Antonio~I Fernandez-Dominguez, Stefan~A Maier, and John~B
  Pendry.
\newblock Harvesting light with transformation optics.
\newblock {\em Science China Information Sciences}, 56(12):1--13, 2013.

\bibitem{pendry2015transforming}
JB~Pendry, Yu~Luo, and Rongkuo Zhao.
\newblock Transforming the optical landscape.
\newblock {\em Science}, 348(6234):521--524, 2015.

\bibitem{aubry2010plasmonic}
Alexandre Aubry, Dang~Yuan Lei, Antonio~I Fern{\'a}ndez-Dom{\'\i}nguez, Yannick
  Sonnefraud, Stefan~A Maier, and John~B Pendry.
\newblock Plasmonic light-harvesting devices over the whole visible spectrum.
\newblock {\em Nano letters}, 10(7):2574--2579, 2010.

\bibitem{fernandez2010collection}
AI~Fern{\'a}ndez-Dom{\'\i}nguez, SA~Maier, and JB~Pendry.
\newblock Collection and concentration of light by touching spheres: a
  transformation optics approach.
\newblock {\em Physical review letters}, 105(26):266807, 2010.

\bibitem{aubry2010broadband}
Alexandre Aubry, Dang~Yuan Lei, Stefan~A Maier, and JB~Pendry.
\newblock Broadband plasmonic device concentrating the energy at the nanoscale:
  The crescent-shaped cylinder.
\newblock {\em Physical Review B}, 82(12):125430, 2010.

\bibitem{fernandez2012theory}
Antonio~I Fernandez-Dominguez, Yu~Luo, Aeneas Wiener, JB~Pendry, and Stefan~A
  Maier.
\newblock Theory of three-dimensional nanocrescent light harvesters.
\newblock {\em Nano letters}, 12(11):5946--5953, 2012.

\bibitem{kraft2014transformation}
Matthias Kraft, JB~Pendry, SA~Maier, and Yu~Luo.
\newblock Transformation optics and hidden symmetries.
\newblock {\em Physical Review B}, 89(24):245125, 2014.

\bibitem{aubry2011plasmonic}
Alexandre Aubry, Dang~Yuan Lei, Stefan~A Maier, and John~B Pendry.
\newblock Plasmonic hybridization between nanowires and a metallic surface: a
  transformation optics approach.
\newblock {\em ACS nano}, 5(4):3293--3308, 2011.

\bibitem{pendry2013capturing}
JB~Pendry, AI~Fern{\'a}ndez-Dom{\'\i}nguez, Yu~Luo, and Rongkuo Zhao.
\newblock Capturing photons with transformation optics.
\newblock {\em Nature Physics}, 9(8):518, 2013.

\bibitem{fernandez2012transformation}
AI~Fern{\'a}ndez-Dom{\'\i}nguez, A~Wiener, FJ~Garc{\'\i}a-Vidal, SA~Maier, and
  JB~Pendry.
\newblock Transformation-optics description of nonlocal effects in plasmonic
  nanostructures.
\newblock {\em Physical review letters}, 108(10):106802, 2012.

\bibitem{fernandez2012insight}
AI~Fern{\'a}ndez-Dom{\'\i}nguez, P~Zhang, Y~Luo, SA~Maier,
  FJ~Garc{\'\i}a-Vidal, and JB~Pendry.
\newblock Transformation-optics insight into nonlocal effects in separated
  nanowires.
\newblock {\em Physical Review B}, 86(24):241110, 2012.

\bibitem{kraft2016transformation}
Matthias Kraft, Yu~Luo, and JB~Pendry.
\newblock Transformation optics: A time-and frequency-domain analysis of
  electron-energy loss spectroscopy.
\newblock {\em Nano letters}, 16(8):5156--5162, 2016.

\bibitem{reddy2019surface}
K~Nireekshan Reddy, Parry~Y Chen, Antonio~I Fern{\'a}ndez-Dom{\'\i}nguez, and
  Yonatan Sivan.
\newblock Surface second-harmonic generation from metallic-nanoparticle
  configurations: A transformation-optics approach.
\newblock {\em Physical Review B}, 99(23):235429, 2019.

\bibitem{zhao2013description}
Rongkuo Zhao, Yu~Luo, AI~Fern{\'a}ndez-Dom{\'\i}nguez, and John~B Pendry.
\newblock Description of van der waals interactions using transformation
  optics.
\newblock {\em Physical review letters}, 111(3):033602, 2013.

\bibitem{luo2014van}
Yu~Luo, Rongkuo Zhao, and John~B Pendry.
\newblock van der waals interactions at the nanoscale: The effects of
  nonlocality.
\newblock {\em Proceedings of the National Academy of Sciences},
  111(52):18422--18427, 2014.

\bibitem{YuPNAS2019}
Sanghyeon Yu and Habib Ammari.
\newblock Hybridization of singular plasmons via transformation optics.
\newblock {\em Proceedings of the National Academy of Sciences},
  116(28):13785--13790, 2019.

\bibitem{holloway2012overview}
Christopher~L Holloway, Edward~F Kuester, Joshua~A Gordon, John O'Hara, Jim
  Booth, and David~R Smith.
\newblock An overview of the theory and applications of metasurfaces: The
  two-dimensional equivalents of metamaterials.
\newblock {\em IEEE Antennas and Propagation Magazine}, 54(2):10--35, 2012.

\bibitem{kildishev2013planar}
Alexander~V Kildishev, Alexandra Boltasseva, and Vladimir~M Shalaev.
\newblock Planar photonics with metasurfaces.
\newblock {\em Science}, 339(6125):1232009, 2013.

\bibitem{Minovich2015}
Alexander~E. Minovich, Andrey~E. Miroshnichenko, Anton~Y. Bykov, Tatiana~V.
  Murzina, Dragomir~N. Neshev, and Yuri~S. Kivshar.
\newblock Functional and nonlinear optical metasurfaces.
\newblock {\em Laser \& Photonics Reviews}, 9(2):195--213, 2015.

\bibitem{GLYBOVSKI20161}
Stanislav~B. Glybovski, Sergei~A. Tretyakov, Pavel~A. Belov, Yuri~S. Kivshar,
  and Constantin~R. Simovski.
\newblock Metasurfaces: From microwaves to visible.
\newblock {\em Physics Reports}, 634:1 -- 72, 2016.
\newblock Metasurfaces: From microwaves to visible.

\bibitem{Monticone2017}
Francesco Monticone and Andrea Al{\`{u}}.
\newblock Metamaterial, plasmonic and nanophotonic devices.
\newblock {\em Reports on Progress in Physics}, 80(3):036401, feb 2017.

\bibitem{SpoofBook}
Paloma~Arroyo Huidobro, Antonio~I. Fern\'andez-Dom\'{i}nguez, John~B. Pendry,
  Luis Martin-Moreno, and Francisco Garcia-Vidal.
\newblock {\em Spoof surface plasmon metamaterials}.
\newblock Cambridge University Press, Cambridge, United Kingdom, 2018.

\bibitem{Ni2011}
Xingjie Ni, Naresh~K Emani, Alexander~V Kildishev, Alexandra Boltasseva, and
  Vladimir~M Shalaev.
\newblock {Broadband Light Bending with Plasmonic Nanoantennas}.
\newblock {\em Science (New York, N.Y.)}, (5):50, 2011.

\bibitem{Yu2011}
Nanfang Yu, Patrice Genevet, Mikhail~A Kats, Francesco Aieta, Jean-Philippe
  Tetienne, Federico Capasso, and Zeno Gaburro.
\newblock {Light propagation with phase discontinuities: generalized laws of
  reflection and refraction.}
\newblock {\em Science (New York)}, 334(6054):333--7, oct 2011.

\bibitem{Genevet2017recent}
Patrice Genevet, Federico Capasso, Francesco Aieta, Mohammadreza
  Khorasaninejad, and Robert Devlin.
\newblock Recent advances in planar optics: from plasmonic to dielectric
  metasurfaces.
\newblock {\em Optica}, 4(1):139--152, Jan 2017.

\bibitem{meinzer2014plasmonic}
Nina Meinzer, William~L Barnes, and Ian~R Hooper.
\newblock Plasmonic meta-atoms and metasurfaces.
\newblock {\em Nature Photonics}, 8(12):889, 2014.

\bibitem{Koppens2011}
Frank H~L Koppens, Darrick~E. Chang, and F.~Javier {Garc{\'{i}}a de Abajo}.
\newblock {Graphene Plasmonics : A Platform for Strong Light-Matter
  Interaction}.
\newblock {\em Nano Lett.}, 11((8)):3370--3377, aug 2011.

\bibitem{Ju2011}
Long Ju, Baisong Geng, Jason Horng, Caglar Girit, Michael Martin, Zhao Hao,
  Hans~A Bechtel, Xiaogan Liang, Alex Zettl, Y~Ron Shen, and Feng Wang.
\newblock {Graphene Plasmonics for Tunable Terahertz Metamaterials.}
\newblock {\em Nat. Nanotechnol.}, 6(10):630--634, oct 2011.

\bibitem{Nikitin2011}
A.~Yu. Nikitin, Francisco Guinea, Francisco~J. Garc{\'{i}}a-Vidal, and Luis
  Mart{\'{i}}n-Moreno.
\newblock {Fields Radiated by a Nanoemitter in a Graphene Sheet}.
\newblock {\em Phys. Rev. B: Condens. Matter Mater. Phys.}, 84(19):195446, nov
  2011.

\bibitem{Grigorenko2012}
a.~N. Grigorenko, M.~Polini, and K.~S. Novoselov.
\newblock {Graphene Plasmonics}.
\newblock {\em Nat. Photonics}, 6(11):749--758, 2012.

\bibitem{Low2014}
Tony Low and Phaedon Avouris.
\newblock {Graphene Plasmonics for Terahertz to Mid-Infrared Applications}.
\newblock {\em ACS Nano}, 8(2):1086--1101, 2014.

\bibitem{Chandezon1980}
J~Chandezon, G~Raoult, and D~Maystre.
\newblock A new theoretical method for diffraction gratings and its numerical
  application.
\newblock {\em Journal of Optics}, 11(4):235--241, jul 1980.

\bibitem{barnes1996}
W.~L. Barnes, T.~W. Preist, S.~C. Kitson, and J.~R. Sambles.
\newblock Physical origin of photonic energy gaps in the propagation of surface
  plasmons on gratings.
\newblock {\em Phys. Rev. B}, 54:6227--6244, Sep 1996.

\bibitem{Kraft2015}
Matthias Kraft, Yu~Luo, S.~A. Maier, and J.~B. Pendry.
\newblock {Designing Plasmonic Gratings with Transformation Optics}.
\newblock {\em Phys. Rev. X}, 5(3):031029, 2015.

\bibitem{Huidobro2017a}
P.~A. Huidobro, Y.~H. Chang, M.~Kraft, and J.~B. Pendry.
\newblock {Hidden symmetries in plasmonic gratings}.
\newblock {\em Phys. Rev. B: Condens. Matter Mater. Phys.}, 95(15):1--8, 2017.

\bibitem{Huidobro2016a}
Paloma~A. Huidobro, Matthias Kraft, Stefan~A. Maier, and John~B. Pendry.
\newblock {Graphene as a Tunable Anisotropic or Isotropic Plasmonic
  Metasurface}.
\newblock {\em ACS Nano}, 10(5):5499--5506, 2016.

\bibitem{PendryPhysRevB2019}
J.~B. Pendry, Paloma~A. Huidobro, and Kun Ding.
\newblock Computing one-dimensional metasurfaces.
\newblock {\em Phys. Rev. B}, 99:085408, Feb 2019.

\bibitem{huidobro2016graphene}
Paloma~A Huidobro, Matthias Kraft, Ren Kun, Stefan~A Maier, and John~B Pendry.
\newblock Graphene, plasmons and transformation optics.
\newblock {\em Journal of Optics}, 18(4):044024, 2016.

\bibitem{peres2013exact}
NMR Peres, Yu~V Bludov, Aires Ferreira, and Mikhail~I Vasilevskiy.
\newblock Exact solution for square-wave grating covered with graphene: surface
  plasmon-polaritons in the terahertz range.
\newblock {\em Journal of Physics: Condensed Matter}, 25(12):125303, 2013.

\bibitem{slipchenko2013}
T~M Slipchenko, M~L Nesterov, L~Martin-Moreno, and a~Yu Nikitin.
\newblock {Analytical Solution for the Diffraction of an Electromagnetic Wave
  by a Graphene Grating}.
\newblock {\em J. Opt. (Bristol, U. K.)}, 15(11):114008, 2013.

\bibitem{chen2017flatland}
Pai-Yen Chen, Christos Argyropoulos, Mohamed Farhat, and J~Sebastian
  Gomez-Diaz.
\newblock Flatland plasmonics and nanophotonics based on graphene and beyond.
\newblock {\em Nanophotonics}, 6(6):1239--1262, 2017.

\bibitem{baudisch2018ultrafast}
Matthias Baudisch, Andrea Marini, Joel~D Cox, Tony Zhu, Francisco Silva,
  Stephan Teichmann, Mathieu Massicotte, Frank Koppens, Leonid~S Levitov,
  F~Javier~Garc{\'\i}a de~Abajo, et~al.
\newblock Ultrafast nonlinear optical response of dirac fermions in graphene.
\newblock {\em Nature communications}, 9(1):1018, 2018.

\bibitem{bokdam2011electrostatic}
Menno Bokdam, Petr~A Khomyakov, Geert Brocks, Zhicheng Zhong, and Paul~J Kelly.
\newblock Electrostatic doping of graphene through ultrathin hexagonal boron
  nitride films.
\newblock {\em Nano letters}, 11(11):4631--4635, 2011.

\bibitem{Fan2015}
Yuancheng Fan, Nian-Hai Shen, Thomas Koschny, and Costas~M. Soukoulis.
\newblock {Tunable Terahertz Meta-Surface with Graphene Cut-Wires}.
\newblock {\em ACS Photonics}, 2(1):151--156, 2015.

\bibitem{Nikitin2012b}
A.~Yu Nikitin, F.~Guinea, F.~J. Garcia-Vidal, and L.~Martin-Moreno.
\newblock {Surface Plasmon Enhanced Absorption and Suppressed Transmission in
  Periodic Arrays of Graphene Ribbons}.
\newblock {\em Phys. Rev. B: Condens. Matter Mater. Phys.}, 85(8):081405(R),
  2012.

\bibitem{vakil2011}
Ashkan Vakil and Nader Engheta.
\newblock {Transformation Optics using Graphene.}
\newblock {\em Science (New York, N.Y.)}, 332(6035):1291--4, jun 2011.

\bibitem{iranzo2018probing}
David~Alcaraz Iranzo, S{\'e}bastien Nanot, Eduardo~JC Dias, Itai Epstein, Cheng
  Peng, Dmitri~K Efetov, Mark~B Lundeberg, Romain Parret, Johann Osmond,
  Jin-Yong Hong, et~al.
\newblock Probing the ultimate plasmon confinement limits with a van der waals
  heterostructure.
\newblock {\em Science}, 360(6386):291--295, 2018.

\bibitem{Huidobro2017}
Paloma {Arroyo Huidobro}, Stefan~A. Maier, and John~B. Pendry.
\newblock {Tunable Plasmonic Metasurface for Perfect Absorption}.
\newblock {\em EPJ Appl. Metamat.}, 4:6, 2017.

\bibitem{Alu2013unbounded}
Nasim~Mohammadi Estakhri and Andrea Al\`u.
\newblock Physics of unbounded, broadband absorption/gain efficiency in
  plasmonic nanoparticles.
\newblock {\em Phys. Rev. B}, 87:205418, May 2013.

\bibitem{Wallen2015anomalous}
Henrik Wallén, Henrik Kettunen, and Ari Sihvola.
\newblock Anomalous absorption, plasmonic resonances, and invisibility of
  radially anisotropic spheres.
\newblock {\em Radio Science}, 50(1):18--28, 2015.

\bibitem{YangPhysRevB2019}
Fan Yang, Yao-Ting Wang, Paloma~A. Huidobro, and John~B. Pendry.
\newblock Nonlocal effects in singular plasmonic metasurfaces.
\newblock {\em Phys. Rev. B}, 99:165423, Apr 2019.

\bibitem{galiffi2018broadband}
Emanuele Galiffi, John~B Pendry, and Paloma~A Huidobro.
\newblock Broadband tunable thz absorption with singular graphene metasurfaces.
\newblock {\em ACS nano}, 12(2):1006--1013, 2018.

\bibitem{sondergaard2012plasmonic}
Thomas S{\o}ndergaard, Sergey~M Novikov, Tobias Holmgaard, Ren{\'e}~L Eriksen,
  Jonas Beermann, Zhanghua Han, Kjeld Pedersen, and Sergey~I Bozhevolnyi.
\newblock Plasmonic black gold by adiabatic nanofocusing and absorption of
  light in ultra-sharp convex grooves.
\newblock {\em Nature communications}, 3:969, 2012.

\bibitem{Pendry2017}
JB~Pendry, Paloma~Arroyo Huidobro, Yu~Luo, and Emanuele Galiffi.
\newblock Compacted dimensions and singular plasmonic surfaces.
\newblock {\em Science}, 358(6365):915--917, 2017.

\bibitem{Yang2018}
Fan Yang, Paloma~A. Huidobro, and J.~B. Pendry.
\newblock Transformation optics approach to singular metasurfaces.
\newblock {\em Phys. Rev. B}, 98:125409, Sep 2018.

\bibitem{Galiffi2019epj}
{Galiffi, Emanuele}, {Pendry, John}, and {Huidobro, Paloma Arroyo}.
\newblock Singular graphene metasurfaces.
\newblock {\em EPJ Appl. Metamat.}, 6:10, 2019.

\bibitem{Benz2016picocavities}
Felix Benz, Mikolaj~K. Schmidt, Alexander Dreismann, Rohit Chikkaraddy, Yao
  Zhang, Angela Demetriadou, Cloudy Carnegie, Hamid Ohadi, Bart de~Nijs, Ruben
  Esteban, Javier Aizpurua, and Jeremy~J. Baumberg.
\newblock Single-molecule optomechanics in
  {\textquotedblleft}picocavities{\textquotedblright}.
\newblock {\em Science}, 354(6313):726--729, 2016.

\bibitem{muljarov2016resonant}
EA~Muljarov and Wolfgang Langbein.
\newblock Resonant-state expansion of dispersive open optical systems: Creating
  gold from sand.
\newblock {\em Physical Review B}, 93(7):075417, 2016.

\bibitem{chen2019generalizing}
Parry~Y Chen, David~J Bergman, and Yonatan Sivan.
\newblock Generalizing normal mode expansion of electromagnetic green’s
  tensor to open systems.
\newblock {\em Physical Review Applied}, 11(4):044018, 2019.

\bibitem{sauvan2013theory}
Christophe Sauvan, Jean-Paul Hugonin, IS~Maksymov, and Philippe Lalanne.
\newblock Theory of the spontaneous optical emission of nanosize photonic and
  plasmon resonators.
\newblock {\em Physical Review Letters}, 110(23):237401, 2013.

\bibitem{kristensen2013modes}
Philip~Tr{\o}st Kristensen and Stephen Hughes.
\newblock Modes and mode volumes of leaky optical cavities and plasmonic
  nanoresonators.
\newblock {\em ACS Photonics}, 1(1):2--10, 2013.

\bibitem{yang2015analytical}
Jianji Yang, Mathias Perrin, and Philippe Lalanne.
\newblock Analytical formalism for the interaction of two-level quantum systems
  with metal nanoresonators.
\newblock {\em Physical Review X}, 5(2):021008, 2015.

\bibitem{hughes2018quantized}
Stephen Hughes, Marten Richter, and Andreas Knorr.
\newblock Quantized pseudomodes for plasmonic cavity qed.
\newblock {\em Optics letters}, 43(8):1834--1837, 2018.

\bibitem{franke2019quantization}
Sebastian Franke, Stephen Hughes, Mohsen~Kamandar Dezfouli, Philip~Tr{\o}st
  Kristensen, Kurt Busch, Andreas Knorr, and Marten Richter.
\newblock Quantization of quasinormal modes for open cavities and plasmonic
  cavity quantum electrodynamics.
\newblock {\em Physical Review Letters}, 122(21):213901, 2019.

\bibitem{dung1998three}
Ho~Trung Dung, Ludwig Kn{\"o}ll, and Dirk-Gunnar Welsch.
\newblock Three-dimensional quantization of the electromagnetic field in
  dispersive and absorbing inhomogeneous dielectrics.
\newblock {\em Physical Review A}, 57(5):3931, 1998.

\bibitem{li2016transformation}
Rui-Qi Li, D~Hern{\'a}ngomez-P{\'e}rez, FJ~Garc{\'\i}a-Vidal, and
  AI~Fern{\'a}ndez-Dom{\'\i}nguez.
\newblock Transformation optics approach to plasmon-exciton strong coupling in
  nanocavities.
\newblock {\em Physical review letters}, 117(10):107401, 2016.

\bibitem{li2018symmetry}
Rui-Qi Li, FJ~Garc{\'\i}a-Vidal, and AI~Fernandez-Dominguez.
\newblock Plasmon-exciton coupling in symmetry-broken nanocavities.
\newblock {\em ACS Photonics}, 5(1):177--185, 2018.

\bibitem{pacheco2016description}
V{\'\i}ctor Pacheco-Peña, Miguel Beruete, Antonio~I
  Fernández-Dom{\'\i}nguez, Yu~Luo, and Miguel Navarro-C{\'\i}a.
\newblock Description of bow-tie nanoantennas excited by localized emitters
  using conformal transformation.
\newblock {\em Acs Photonics}, 3(7):1223--1232, 2016.

\bibitem{pacheco2017aluminum}
V{\'\i}ctor Pacheco-Pe{\~n}a, Antonio~I Fern{\'a}ndez-Dom{\'\i}nguez, Yu~Luo,
  Miguel Beruete, and Miguel Navarro-C{\'\i}a.
\newblock Aluminum nanotripods for light-matter coupling robust to nanoemitter
  orientation.
\newblock {\em Laser \& Photonics Reviews}, 11(5):1700051, 2017.

\bibitem{cuartero2018light}
A~Cuartero-Gonz{\'a}lez and AI~Fern{\'a}ndez-Dom{\'\i}nguez.
\newblock Light-forbidden transitions in plasmon-emitter interactions beyond
  the weak coupling regime.
\newblock {\em ACS Photonics}, 5(8):3415--3420, 2018.

\bibitem{cuartero2019dipolar}
A~Cuartero-Gonz{\'a}lez and AI~Fern{\'a}ndez-Dom{\'\i}nguez.
\newblock Dipolar and quadrupolar excitons coupled to a
  nanoparticle-on-a-mirror cavity.
\newblock {\em arXiv preprint arXiv:1905.09893}, 2019.

\bibitem{breuer2002theory}
Heinz-Peter Breuer, Francesco Petruccione, et~al.
\newblock {\em The theory of open quantum systems}.
\newblock Oxford University Press on Demand, 2002.

\bibitem{gonzalez2014reversible}
Alejandro Gonz{\'a}lez-Tudela, PA~Huidobro, Luis Mart{\'\i}n-Moreno, C~Tejedor,
  and FJ~Garc{\'\i}a-Vidal.
\newblock Reversible dynamics of single quantum emitters near metal-dielectric
  interfaces.
\newblock {\em Physical Review B}, 89(4):041402, 2014.

\bibitem{delga2014quantum}
A~Delga, J~Feist, J~Bravo-Abad, and FJ~Garcia-Vidal.
\newblock Quantum emitters near a metal nanoparticle: strong coupling and
  quenching.
\newblock {\em Physical review letters}, 112(25):253601, 2014.

\bibitem{giannini2011plasmonic}
Vincenzo Giannini, Antonio~I Fern{\'a}ndez-Dom{\'\i}nguez, Susannah~C Heck, and
  Stefan~A Maier.
\newblock Plasmonic nanoantennas: fundamentals and their use in controlling the
  radiative properties of nanoemitters.
\newblock {\em Chemical reviews}, 111(6):3888--3912, 2011.

\bibitem{demetriadou2017spatiotemporal}
Angela Demetriadou, Joachim~M Hamm, Yu~Luo, John~B Pendry, Jeremy~J Baumberg,
  and Ortwin Hess.
\newblock Spatiotemporal dynamics and control of strong coupling in plasmonic
  nanocavities.
\newblock {\em ACS Photonics}, 4(10):2410--2418, 2017.

\bibitem{esteban2014strong}
Ruben Esteban, Javier Aizpurua, and Garnett~W Bryant.
\newblock Strong coupling of single emitters interacting with phononic infrared
  antennae.
\newblock {\em New Journal of Physics}, 16(1):013052, 2014.

\bibitem{saez2017enhancing}
Roc{\'\i}o S{\'a}ez-Bl{\'a}zquez, Johannes Feist,
  AI~Fern{\'a}ndez-Dom{\'\i}nguez, and FJ~Garc{\'\i}a-Vidal.
\newblock Enhancing photon correlations through plasmonic strong coupling.
\newblock {\em Optica}, 4(11):1363--1367, 2017.

\bibitem{laussy2008strong}
Fabrice~P Laussy, Elena Del~Valle, and Carlos Tejedor.
\newblock Strong coupling of quantum dots in microcavities.
\newblock {\em Physical review letters}, 101(8):083601, 2008.

\bibitem{saez2018photon}
R~S{\'a}ez-Bl{\'a}zquez, J~Feist, FJ~Garc{\'\i}a-Vidal, and
  AI~Fern{\'a}ndez-Dom{\'\i}nguez.
\newblock Photon statistics in collective strong coupling: Nanocavities and
  microcavities.
\newblock {\em Physical Review A}, 98(1):013839, 2018.

\end{thebibliography}
\bibliographystyle{unsrt}

\end{document}